\documentclass[preprint,showpacs,nofootinbib,eqsecnum,amsfonts,amsmath,aps]{revtex4}

\allowdisplaybreaks
\newtheorem{theorem}{Theorem}

\begin{document}

\title{Time-symmetric initial data for binary black holes\\ in
numerical relativity} \author{Luc Blanchet}\email{blanchet@iap.fr}
\affiliation{Gravitation et Cosmologie (GReCO),\\ Institut
d'Astrophysique de Paris --- C.N.R.S.,\\ 98\textsuperscript{~\!bis}
boulevard Arago, 75014 Paris, France}\affiliation{Faculty of Science
and Technology,\\ Hirosaki University, Hirosaki 036-8561, Japan}
\date{\today}

\begin{abstract}
We look for physically realistic initial data in numerical relativity
which are in agreement with post-Newtonian approximations. We propose
a particular solution of the time-symmetric constraint equation,
appropriate to two momentarily static black holes, in the form of a
conformal decomposition of the spatial metric. This solution is
isometric to the post-Newtonian metric up to the 2PN order. It
represents a non-linear deformation of the solution of Brill and
Lindquist, i.e. an asymptotically flat region is connected to two
asymptotically flat (in a certain weak sense) sheets, that are the
images of the two singularities through appropriate inversion
transformations. The total ADM mass $M$ as well as the individual
masses $m_1$ and $m_2$ (when they exist) are computed by surface
integrals performed at infinity. Using second order perturbation
theory on the Brill-Lindquist background, we prove that the binary's
interacting mass-energy $M-m_1-m_2$ is well-defined at the 2PN order
and in agreement with the known post-Newtonian result.
\end{abstract}


\maketitle

\section{Motivation and relation to other works}\label{sec1}

The numerical computation of the collision of two black holes is of
paramount importance for the observation of gravitational waves by the
network of laser-interferometric detectors. When investigating this
problem the ten Einstein field equations are separated into: (i) four
constraint equations that are to be satisfied by some initial data
given on an initial three-dimensional Cauchy hypersurface; (ii) six
hyperbolic-like equations describing the dynamical evolution of the
gravitational field on neighbouring hypersurfaces. The Bianchi
identities guarantee that the constraint equations are satisfied on
neighbouring hypersurfaces if they are on the initial
hypersurface. There are infinitely many ways that the initial data can
be chosen to represent the starting state of the evolution of black
holes. It is widely admitted that the problem of choosing physically
realistic initial conditions for the collision of two black holes has
not yet been solved. There has been a lot of concern in the
litterature \cite{cook} for knowing what would really motivate
physically a particular choice of initial data.

Let us consider the problem of {\it time-symmetric} initial data,
which are physically appropriate to two momentarily static black
holes, i.e. with zero initial velocities. The dynamical evolution of
time-symmetric data describes the subsequent {\it head-on} collision
of the two black holes. In this situation the constraint equations
reduce to the Hamiltonian or scalar contraint equation $R=0$ (in
vacuum --- the case appropriate to black holes), with $R$ being the
three-dimensional scalar curvature. Considering as usual
\cite{lichne,brill,york} a conformal decomposition of the spatial
metric (spatial indices $i,j,\cdots=1,2,3$),

\begin{equation}\label{1}
\gamma_{ij}=\Psi^4\widetilde{\gamma}_{ij}\;,
\end{equation}
where $\gamma_{ij}$ is the physical metric and $\widetilde{\gamma}_{ij}$
denotes the conformal (unphysical) metric, we obtain the Lichnerowicz
\cite{lichne} equation, which is an elliptic-type equation to be
satisfied by the conformal factor $\Psi$. In the time-symmetric case
that equation becomes

\begin{equation}\label{2}
\widetilde{\Delta}\Psi=\frac{\Psi}{8}\widetilde{R}\;.
\end{equation}
The scalar curvature $\widetilde{R}$ and Laplacian
$\widetilde{\Delta}$ are the ones associated with the conformal metric
$\widetilde{\gamma}_{ij}$. The interest of the conformal decomposition
is that by solving Eq. (\ref{2}) we generate a physical solution
$\gamma_{ij}$ of the constraint equation starting from {\it any}
choice for the conformal metric $\widetilde{\gamma}_{ij}$ ({\it a
priori}). Therefore the problem of initial conditions resides in the
choice of a physically well-motivated conformal metric
$\widetilde{\gamma}_{ij}$.

The simplest choice of initial conditions (one motivated by simplicity
rather than by physics) is the one for which $\widetilde{\gamma}_{ij}$
equals the flat metric $\delta_{ij}$. In that case Eq. (\ref{2})
reduces to the flat-space Laplace equation. Some exact solutions,
appropriate to (momentarily static) black holes, have been obtained by
Misner \cite{misner} and Lindquist \cite{lindquist}, and Brill and
Lindquist \cite{BL}. The solution of Brill and Lindquist [for which
the conformal factor $\Psi^{\rm{BL}}$ takes the form of
Eq. (\ref{10b}) below] is particularly interesting: it describes the
``geometrostatics'' of two black holes, consisting of three
asymptotically flat regions connected by two Einstein-Rosen bridges
(actually, the solution is known for $N$ black holes). The one region
containing the two throats is supposed to represent our universe,
while the two sheets expanding behind it are associated with the two
black holes. The beauty of the Brill-Lindquist solution is that not
only the total ADM mass-energy $M$ of the binary, but also the two
individual masses $m_1$ and $m_2$ of the black holes, are computed
``at infinity''. One can use for instance standard surface integrals
extending on topological two-spheres at infinity. The binary's
geometrostatic energy (i.e. the gravitational interacting or binding
energy, in the center-of-mass frame) is therefore computed
unambiguously as $\frac{E}{c^2}=M-m_1-m_2$.

Yet the solution of Brill and Lindquist, despite its undeniable
interest, is not ``physically realistic'' in the sense that it
differs, in the limit $c\to +\infty$, from the three-metric found for
a post-Newtonian solution. Indeed the post-Newtonian metric generated
by two point-particles is known to deviate from conformal flatness at
the 2PN order (see e.g. Ref. \cite{BFP98})\,\footnote{In this paper
$c$ and $G$ denote the speed of light and the gravitational
constant. As usual the $n$PN order means the terms of order
$\frac{1}{c^{2n}}$ when $c\to +\infty$.}. In consequence both the
Brill-Lindquist metric and the associated binding energy $E$ disagree
with the post-Newtonian results from the 2PN order.

A compelling motivation for constructing physically realistic initial
data is the agreement with the post-Newtonian approximation when $c\to
+\infty$. Recall that the post-Newtonian theory provides some explicit
expressions for the metric, equations of motion and energy of binary
systems of point-particles. The post-Newtonian metric is valid in the
binary's ``near-zone'' (size of near-zone is much less than a
gravitational wavelength), but has been proved to come from the
re-expansion when $c\to +\infty$ of a ``global''
(post-Minkowskian-type) solution, defined everywhere in space-time
including the wave zone \cite{livrev}. Furthermore, in the
post-Newtonian approach the modelling of black holes by point-like
particles --- i.e. technically by Dirac delta-functions in the
stress-energy tensor --- is rather well justified. We shall provide
below some further evidence that the ``post-Newtonian masses'' of
point-particles are indeed identical to the black-hole
masses\,\footnote{See also, in a similar context, the matching of a
1PN solution for the orbital motion of point-particles to two
perturbed Schwarzschild black holes \cite{alvi}.}.

It has been suggested \cite{diener} (see also \cite{BCLT}) that, in
order of taking into account the post-Newtonian physics into the
initial data, one should adopt for the {\it conformal} metric
$\widetilde{\gamma}_{ij}$ directly a post-Newtonian solution. In such
a proposal one expects that by correcting the post-Newtonian solution
by means of a conformal factor $\Psi^4$ (computed numerically), one
will somehow be able to ``compensate'' in the physical metric
$\gamma_{ij}$ (i.e., in fact, to cancel out) the systematic
higher-order post-Newtonian error terms that are neglected in
$\widetilde{\gamma}_{ij}$.

There has been other proposals for realistic initial data, in
particular built on the relaxation of the assumption of conformal
flatness \cite{cook}. One of these is to use for
$\widetilde{\gamma}_{ij}$ a linear combination of two (boosted
versions of) Kerr-Schild metrics \cite{matzner,BMNM}. It is not known
if the physical metric which is generated in this way from the
numerically-computed conformal factor, is consistent with
post-Newtonian (e.g. 2PN) calculations.

On the other hand, we should remark that to which extent the
hypothesis of conformal flatness introduces some unphysical spurious
(and numerically important) effects remains an issue. Indeed, relaxing
this hypothesis may not always be a panacea. A quite different idea
for settling the initial conditions of two black holes is to solve
numerically a subset of the Einstein field equations (the four
constraint equations plus one evolution equation), under the two
premices of conformal flatness and the existence of a helical Killing
symmetry \cite{GGB1,GGB2}. Despite the conformal flatness of the
spatial metric, the latter calculation gives results in good agreement
with post-Newtonian predictions
\cite{GGB2,B02ico,DGG02}. Nevertheless, we think that it is important,
at some stage, to get rid of the hypothesis of conformal flatness.

We emphasize furthermore that the agreement between numerical
relativity \cite{GGB1,GGB2} and post-Newtonian theory
\cite{GGB2,B02ico,DGG02} holds up to the very relativistic regime of
the innermost circular orbit (ICO), where the orbital velocities are
of the order of 50\% of the speed of light\,\footnote{The ICO is
defined by the minimum of the binary's energy function for circular
orbits. It represents a useful reference point for the numerical
\cite{GGB1,GGB2} and 3PN \cite{B02ico} calculations because for both
of them the ICO is well-defined and can be meaningfully
compared. However, the radiation reaction terms are neglected in its
definition, so the ICO probably does not have a rigorous physical
meaning in a context of exact radiative solutions.}. In
Refs. \cite{B02ico} it is suggested that the result for the ICO of two
black holes with comparable masses, at the 3PN
approximation\,\footnote{We mean the standard Taylor-expanded form of
the approximation --- without using any post-Newtonian resummation
techniques.}, is likely to be close to the ``exact'' solution, within
1\% of fractional accuracy or better. This constitutes a motivation
for advocating that the black-hole initial conditions, which are to be
set (presumably) around the location of the ICO, should be in
agreement with post-Newtonian theory.

In the present paper we propose an alternative way for incorporating
the post-Newtonian information into the initial data of black-hole
binaries (in the time-symmetric case). We find, in Section \ref{sec2},
a simple expression for the conformal metric
$\widetilde{\gamma}_{ij}$, which is such that the corresponding {\it
physical} metric $\gamma_{ij}$ is isometric (i.e. differs by a
coordinate transformation) to the standard post-Newtonian spatial
metric at the 2PN order in the limit $c\to +\infty$. At the same time,
the solution is defined globally in space, with a global structure
similar to the one of Brill and Lindquist. Our solution is not
``exact'', but exists as a certain non-linear perturbation,
investigated in Section \ref{sec3}, of the Brill-Lindquist solution,
playing here the role of a ``background'' metric.  Most importantly,
in Section \ref{sec4} we investigate the asymptotic structure of the
solution, and compute ``geometrically'' the masses $M$ and $m_1$,
$m_2$, i.e. by surface integrals performed in their respective domains
at infinity. In Section \ref{sec5} the binary's interacting energy,
deduced from the previous masses, is proved to be in agreement with
the known post-Newtonian energy up to the 2PN order. (We shall find,
however, that the definition we adopt for the two individual masses
$m_1$ and $m_2$ makes sense only up to the 2PN order.)

\section{Conformal decomposition of the spatial metric}\label{sec2}

\subsection{Definition of the conformal metric}\label{sec2A}

The conformal metric we propose, appropriate to two black holes of
masses $m_1$ and $m_2$ (to agree later with the ``geometrical''
masses), located at the singular points ${\bf y}_1$ and ${\bf y}_2$,
and momentarily at rest (${\bf v}_1={\bf v}_2={\bf 0}$), is

\begin{equation}\label{3}
\widetilde{\gamma}_{ij} = \delta_{ij}
-\frac{8G^2m_1m_2}{c^4}\frac{\partial^2g} {\partial y_1^{<i}\partial
y_2^{j>}}\;.
\end{equation}
The function $g$ introduced here represents an elementary ``kernel''
playing an important role in post-Newtonian calculations
\cite{S87,DI91,BDI95,WWi96,BFP98,AF97}. It depends on the ``field''
point ${\bf x}$ on the one hand, and on the pair of ``source'' points
${\bf y}_1,\,{\bf y}_2$ on the other hand; it is defined by

\begin{equation}\label{4}
g({\bf x};{\bf y}_1,{\bf y}_2)=\ln\left(r_1+r_2+r_{12}\right)\;,
\end{equation}
where $r_1=|{\bf x}-{\bf y}_1|$ and $r_2=|{\bf x}-{\bf y}_2|$ are the
distances to the black holes, and $r_{12}=|{\bf y}_1-{\bf y}_2|$ is
their separation. The function $g$ satisfies, in the sense of
distributions (i.e. for all values of ${\bf x}$, including the singular
points ${\bf y}_1$ and ${\bf y}_2$),

\begin{equation}\label{5}
\Delta g=\frac{1}{r_1r_2}\;,
\end{equation}
where $\Delta$ denotes the usual flat-space Laplacian with respect to
the field point ${\bf x}$. In Eq. (\ref{3}) the derivatives are taken
with respect to the two source points\,\footnote{The following
explicit formula holds:
$${}_ig_j\equiv\frac{\partial^2g}{\partial y_1^{i}\partial y_2^{j}}=
\frac{n_{12}^in_{12}^j-\delta^{ij}}{r_{12}(r_1+r_2+r_{12})}+\frac{(n_{12}^i-
n_1^i)(n_{12}^j+n_2^j)}{(r_1+r_2+r_{12})^2}\;.$$ Here,
$n_1^i=(x^i-y_1^i)/r_1$, $n_2^i=(x^i-y_2^i)/r_2$ and
$n_{12}^i=(y_1^i-y_2^i)/r_{12}$. See Ref. \cite{BFP98} for further
discussion and formulas concerning the function $g$.}. The carets
around the indices refer to the symmetric and trace-free (STF)
projection: $T_{<ij>}\equiv\frac{1}{2}\left(
T_{ij}+T_{ji}\right)-\frac{1}{3}\delta_{ij}T_{kk}$; so the trace of
the metric (\ref{3}) is normalized to be
$\widetilde{\gamma}_{ii}=3$. It is of course quite natural (and useful
in practice) to impose that the deviation of the conformal metric from
flat space be trace-free.

Our proposal is to generate, by means of numerical techniques
(i.e. with the help of elliptic solvers), a conformal factor $\Psi$
solving the constraint equation (\ref{2}) corresponding to the
particular choice of conformal metric (\ref{3}). The metric
$\gamma_{ij}=\Psi^4\widetilde{\gamma}_{ij}$ we obtain in this way will
incorporate the post-Newtonian (2PN) physics of the initial state of
the head-on collision of two black holes\,\footnote{See
Refs. \cite{Smarr,AHSSS} for numerical calculations of the head-on
collision of black holes; see also Ref. \cite{SPW95} for a
post-Newtonian calculation.}. The latter assertion will now be proved,
for the rest of the paper, with the help of analytic perturbation
methods.

\subsection{Relation with the post-Newtonian metric}\label{sec2B}

The first result shows that the ``near-zone'' behaviour of the
solution (i.e. $r/c\to 0$) is physically sound.

\begin{theorem}\label{th1}
The post-Newtonian expansion (when $c\to +\infty$) of the solution
$\gamma_{ij}=\Psi^4\widetilde{\gamma}_{ij}$ of the constraint equation
${\rm (\ref{2})}$, where $\widetilde{\gamma}_{ij}$ is defined by
Eq. ${\rm (\ref{3})}$, differs from the standard post-Newtonian
spatial metric, calculated by standard post-Newtonian methods, by a
mere change of coordinates at the ${\rm 2PN}$ order, i.e.

\begin{equation}\label{6}
\gamma_{ij} = {{\rm g}_{ij}^{\rm 2PN}}_{\bigl|_{{\bf v}_1={\bf
v}_2={\bf 0}}}+\partial_i\xi_j+\partial_j\xi_i+{\cal
O}\left(\frac{1}{c^6}\right)\;.
\end{equation}
\end{theorem}

To be precise, by ${\rm g}_{ij}^{\rm 2PN}$ we mean the spatial metric
in {\it harmonic coordinates}, when truncated at the 2PN order, that
is given by Eq. (7.2c) in Ref. \cite{BFP98}. As indicated in
Eq. (\ref{6}), we must set the particles' velocities ${\bf v}_1$ and
${\bf v}_2$ to zero in the post-Newtonian metric in order to conform
with the assumption of time-symmetry. The remainder ${\cal
O}\left(\frac{1}{c^6}\right)$ stands for the neglected 3PN and
higher-order terms. 

The proof of Theorem \ref{th1} is easily achieved on the basis of a
post-Newtonian iteration of Eq. (\ref{2}). At the 2PN order, this
equation becomes

\begin{equation}\label{6'}
\Delta\Psi =
-\frac{G^2m_1m_2}{c^4}\partial_{ij}\left({}_{<i}g_{j>}\right)+{\cal
O}\left(\frac{1}{c^6}\right)\;,
\end{equation}
where we denote ${}_ig_j\equiv\frac{\partial^2g}{\partial
y_1^{i}\partial y_2^{j}}$. The most general solution reads as

\begin{equation}\label{6''}
\Psi = \psi -\frac{G^2m_1m_2}{2c^4}{\rm D}\left(\frac{g}{3}+\frac{r_1
+r_2}{2r_{12}}\right)+{\cal O}\left(\frac{1}{c^6}\right)\;,
\end{equation}
where $\psi$ represents a solution of the homogeneous equation
(i.e. $\Delta\psi=0$)\,\footnote{We employ the notational shorthand
${\rm D}\equiv\frac{\partial^2}{\partial y_1^i\partial y_2^i}$ (so
that ${\rm D}g={}_ig_i$). Notice the useful relations
\begin{eqnarray*}
{\rm D}g&=&\frac{1}{2r_1r_2}-\frac{1}{2r_1r_{12}}
-\frac{1}{2r_2r_{12}}\;,\\ \partial_{ij}\left({}_ig_j\right)&=&{\rm
D}\left[\frac{1}{2r_1r_2}+\frac{1}{2r_1r_{12}}+\frac{1}{2r_2r_{12}}\right]\;.
\end{eqnarray*}}. By comparing with the post-Newtonian metric (see
Eq. (7.2c) in Ref. \cite{BFP98} in which ${\bf v}_1={\bf v}_2={\bf
0}$), we readily find that the latter solution is uniquely specified,
at the 2PN order, as being

\begin{equation}\label{6'''}
\psi = 1 + \frac{G m_1}{2c^2r_1}\left(1-\frac{G
m_2}{2c^2r_{12}}\right) +\frac{G m_2}{2c^2r_2}\left(1-\frac{G
m_1}{2c^2r_{12}}\right) + {\cal O}\left(\frac{1}{c^6}\right)\;.
\end{equation}
Also uniquely determined is the expression of the vector $\xi_i$ in
Eq. (\ref{6}), which represents an infinitesimal gauge
transformation. At the 2PN order it takes the expression

\begin{equation}\label{7}
\xi_i=\frac{G^2m_1^2}{4c^4}\frac{n_1^i}{r_1}
+\frac{G^2m_2^2}{4c^4}\frac{n_2^i}{r_2}+{\cal
O}\left(\frac{1}{c^6}\right)\;,
\end{equation}
where ${\bf n}_1\equiv ({\bf x}-{\bf y}_1)/r_1$ and ${\bf n}_2\equiv
({\bf x}-{\bf y}_2)/r_2$. The system of spatial coordinates employed
in the present paper, i.e. which corresponds to a conformal metric of
the particular form displayed by Eq. (\ref{3}), is thereby determined,
and related up to the 2PN order to the harmonic coordinate system used
in Ref. \cite{BFP98}, by the coordinate change $\delta x^i=\xi^i$
(with $\xi^i=\delta^{ij}\xi_j$).

Summarizing this section, we have found a conformal decomposition of
the metric $\gamma_{ij}$ that is in agreement with the post-Newtonian
metric up to the 2PN order. Notice, however, that the conformal metric
we propose is not unique, because we can always add higher-order terms
(e.g. 3PN) to Eq. (\ref{3}). {\it A contrario}, this means that with
the metric (\ref{3}) one should not expect Theorem \ref{th1} to hold
at the 3PN order and beyond. At the 3PN order for instance the
conformal metric will be more complicated than the simple expression
(\ref{3}).

\section{Perturbation of the Brill-Lindquist solution}\label{sec3}

Having checked the near-zone structure of the metric, let us
investigate some of its {\it global} properties, when it is viewed as
a solution of the constraint equation (\ref{2}) that is {\it a priori}
valid everywhere on a space-like hypersurface. For this purpose, we
impose that $\gamma_{ij}$ is a non-linear perturbation of the exact
solution of Brill and Lindquist \cite{BL} --- which is conformally
flat ($\widetilde{\gamma}_{ij}^{\rm{BL}} = \delta_{ij}$). This will
mean that the topology of our solution is identical to the topology of
the Brill-Lindquist solution, i.e. be ``three-sheeted'', in contrast
with the two-sheeted topology of the Misner-Lindquist solution
\cite{misner,lindquist}.

\subsection{Hierarchy of perturbation equations}\label{sec3A}

Let us write Eq. (\ref{3}) in a more transparent form,
$\widetilde{\gamma}_{ij} = \delta_{ij}+\beta s_{ij}$, where

\begin{subequations}\label{9}\begin{eqnarray}
\beta&\equiv&-\frac{8G^2m_1m_2}{c^4}\;,\label{9a}\\s_{ij}&\equiv&
\frac{\partial^2g}{\partial y_1^{<i}\partial
y_2^{j>}}\quad\hbox{(such that $s_{ii}=0$)}\;.\label{9b}
\end{eqnarray}\end{subequations}
It is helpful to view the non-linear perturbation we want to consider,
as being generated by the ``seed'' or ``generating'' function
$s_{ij}$, and to interpret the parameter $\beta$ as the magnitude of
that perturbation. In the following, $\beta$ will play the role of a
``book-keeping'' parameter allowing us to label the successive
non-linear perturbation orders. Besides the non-conformally flat piece
of the metric brought about by $s_{ij}$, it is evident that we must
also introduce a perturbation in the conformal factor. We pose

\begin{subequations}\label{10}\begin{eqnarray}
\Psi&=&\psi+\sigma\;,\label{10a}\\
\psi&=&1+\frac{\alpha_1}{r_1}+\frac{\alpha_2}{r_2}\;,\label{10b}
\end{eqnarray}\end{subequations}
where $\psi$ denotes the Brill-Lindquist conformal factor \cite{BL},
parametrized by two constants $\alpha_1$ and $\alpha_2$ (in the case
of two particles). The quantity $\sigma$ denotes a certain
perturbation of $\psi$.

With full generality --- within the present perturbative framework
---, we look for the expression of $\sigma$ in the form of an infinite
power series in $\beta$, solving the equation (\ref{2}):

\begin{equation}\label{11}
\sigma=\sum_{n=1}^{+\infty}\beta^n\sigma_{(n)}\;.
\end{equation}
We insist that in our approach the Brill-Lindquist solution plays the
role of the {\it background} metric (it depends solely on $\alpha_1$
and $\alpha_2$), while the function $s_{ij}$ yields a non-linear
deformation of this background, perturbatively ordered by the
book-keeping parameter $\beta$. For a given choice of $s_{ij}$, we
expect that the resulting solution is unique (at least in a sense of
formal power series in $\beta$). Because $\beta$ involves two mass
factors $m_1$ and $m_2$, so it is proportional to $G^2$, the
perturbation series we look for is like a ``double'' post-Minkowskian
expansion, going ``twice as fast'' as the usual post-Minkowskian
expansion when $G\to 0$. Considering this series on the point of view
of a post-Newtonian re-expansion, we see that each non-linear order
brings in a new factor $1/c^4$, so that our perturbation series can be
said to go by --- quite efficient indeed --- steps of 2PN orders.

Let us compare our definitions (\ref{10})-(\ref{11}) with the result
(\ref{6''})-(\ref{6'''}) provided by the agreement with the 2PN
metric. We find that the constants $\alpha_1,\alpha_2$ are determined
with relative 1PN accuracy,

\begin{equation}\label{11'}
\alpha_1=\frac{G m_1}{2c^2}\left[1-\frac{G m_2}{2r_{12}c^2}+{\cal
O}\left(\frac{1}{c^4}\right)\right]\quad\hbox{and $~1\leftrightarrow
2$}\;,
\end{equation}
and that with this accuracy they agree with the prediction of the
Brill-Lindquist solution\,\footnote{We recall that $\alpha_1^{\rm BL}$
and $\alpha_2^{\rm BL}$ in the case of the Brill-Lindquist solution
are related to the masses by the exact relations
\begin{eqnarray*}
\alpha_1^{\rm BL}\left[1+\frac{\alpha_2^{\rm
BL}}{r_{12}}\right]&=&\frac{G m_1}{2c^2}\quad\hbox{and
$~1\leftrightarrow 2$}\;,\\ \alpha_1^{\rm BL}+\alpha_2^{\rm
BL}&=&\frac{G M}{2c^2}\;.
\end{eqnarray*}}. In addition, we find that the perturbation $\sigma$ in the
conformal factor is given by the second term in the right side of
Eq. (\ref{6''}), which comes in only at the 2PN order --- it is purely
linear in $\beta$. Hence,

\begin{equation}\label{11''}
\sigma_{(1)}=\frac{1}{16}{\rm D}\left(\frac{g}{3}
+\frac{r_1+r_2}{2r_{12}}\right)+{\cal
O}\left(\frac{1}{c^2}\right)\;,
\end{equation}
with all higher-order $\sigma_{(n)}$'s being negligible with this
approximation.  Actually, Eqs. (\ref{11'})-(\ref{11''}) correspond to
a particular ``sharing'' of the terms between $\alpha_1,\alpha_2$ on
the one hand, and $\sigma_{(1)}$ on the other hand, since we can
always add to $\sigma_{(1)}$ some ``homogeneous'' terms $\sim 1/r_1$
and $\sim 1/r_2$ without changing the equation for $\sigma_{(1)}$. It
is obvious that such a sharing is physically irrelevant. However we
shall forbid a different sharing of terms by adopting below the
prescription (which represents simply a convenient choice) that
$\sigma_{(1)}$ and all subsequent iterations solve the equation for
the conformal factor in the sense of distributions. Some results more
complete than (\ref{11'})-(\ref{11''}) will be obtained below.

By inserting the perturbation ansatz (\ref{11}) into the constraint
equation (\ref{2}), and by identifying each of the coefficients of the
successive powers of $\beta$ in both sides of the equation, we obtain
a hierarchy (indexed by $n\in {\mathbb N}$) of Poisson-type equations:

\begin{equation}\label{12}
\Delta\sigma_{(n)}=\Sigma_{(n)}\left[\sigma_{(1)},\cdots,
\sigma_{(n-1)}\right]\;,
\end{equation}
where we recall that $\Delta\equiv\partial_i\partial_i$. The
$n$th-order source term $\Sigma_{(n)}$ depends on the solutions of the
preceding iterations $\sigma_{(1)}, \cdots,\sigma_{(n-1)}$, on the
``generating'' function $s_{ij}$ and on the ``background'' conformal
factor $\psi$. Notice that $\psi$ satisfies the equation
$\Delta\psi=-4\pi(\alpha_1\delta_1+\alpha_2\delta_2)$, where e.g.
$\delta_1\equiv\delta({\bf x}-{\bf y}_1)$ denotes the Dirac function
at the point ${\bf y}_1$. Obviously we may include the equation for
the background conformal factor into our hierarchy of equations by
posing $\Psi=\sum_{n=0}^{+\infty}\beta^n\sigma_{(n)}$, with
$\sigma_{(0)}=\psi$ and
$\Sigma_{(0)}=-4\pi(\alpha_1\delta_1+\alpha_2\delta_2)$.

\subsection{Analytic closed form of the linearized solution}\label{sec3B}

To the linearized order the perturbation equation reads explicitly

\begin{equation}\label{13}
\Delta\sigma_{(1)}=\frac{1}{8}\psi\partial_{ij}s_{ij}
+\partial_js_{ij}\partial_i\psi +s_{ij}\partial_{ij}\psi\;.
\end{equation}
Interestingly, this equation turns out to be solvable in analytic
closed form. We find that the {\it unique} solution of Eq. (\ref{13}),
that is valid in the sense of distributions and tends to zero at
spatial infinity (i.e. when $r\equiv |{\bf x}|\to +\infty$), is

\begin{eqnarray}\label{13'}
\sigma_{(1)}&=&\frac{1}{16}{\rm D}\left(\frac{g}{3}
+\frac{r_1+r_2}{2r_{12}}\right)\nonumber\\
&+&\alpha_1\left\{\frac{H_1}{2}+\frac{K_1}{32}-\frac{1}{32}{\rm
D}\left(\frac{1}{r_2}\ln\Big[\frac{r_1}{r_{12}}\Big]\right)+\frac{9}{32}{\rm
D}\left(\frac{\ln r_1}{r_{12}}\right)+\frac{{\rm D}g}{12r_1}\right.
\nonumber\\ &&\quad \left.+
\frac{1}{4}\Delta_1\left(\frac{g}{r_{12}}\right)
-\frac{1}{32}\Delta_2\left(\frac{g}{r_{12}}\right)
-\frac{1}{24r_1r_{12}^2}+\frac{1}{32r_2r_{12}^2}\right\}\nonumber\\
&+&\alpha_2\Big\{1\leftrightarrow 2\Big\}\;.
\end{eqnarray}
Besides the already met shorthand ${\rm
D}\equiv\frac{\partial^2}{\partial y_1^i\partial y_2^i}$, we denote
the Laplacians with respect to the source points by
$\Delta_1\equiv\frac{\partial^2}{\partial y_1^i\partial y_1^i}$ and
$\Delta_2\equiv\frac{\partial^2}{\partial y_2^i\partial y_2^i}$. In
Eq. (\ref{13'}), the two terms proportional to $\alpha_1$ and
$\alpha_2$ are deduced from each other by label exchange
$1\leftrightarrow 2$. The solution involves the special functions
$H_1$ and $K_1$ (and $1\leftrightarrow 2$) which were introduced in
Refs. \cite{BDI95,BFP98} for solving some elementary equations (in the
sense of distributions) in the problems of equations of motion and
wave generation at the 2PN order. We have,

\begin{subequations}\label{13''}\begin{eqnarray}
\Delta H_1 &=& 2\,{}_ig_j\,\partial_{ij}\left(\frac{1}{r_1}\right)\;,\\
\Delta K_1 &=& 2\,{\rm D}^2\left(\frac{\ln r_1}{r_2}\right)\;,
\end{eqnarray}\end{subequations}
in which ${}_ig_j\equiv\frac{\partial^2g}{\partial y_1^i\partial
y_2^j}$. The functions $H_1$ and $K_1$ admit the closed-form
expressions given by Eqs. (3.48)-(3.51) in Ref. \cite{BDI95}, or
equivalently by Eqs. (6.3)-(6.5) in Ref. \cite{BFP98}; for the present
purpose we adopt the formulas\,\footnote{We notice here (since it was
not noticed in Refs. \cite{BDI95,BFP98}) that $K_1$ can also be given
an interesting form in connection with a simple elementary kernel
$k_1$, {\it viz}
\begin{eqnarray*}
\Delta k_1 &=& \frac{1}{r_1^2r_2}\;,\\ K_1 &=& {\rm
D}\left(\frac{1}{r_2}\ln\Big[\frac{r_1}{r_{12}}\Big]-k_1\right)\;.
\end{eqnarray*}}

\begin{subequations}\label{13'''}\begin{eqnarray}
H_1 &=& \Delta_1\left[\frac{g}{2r_{12}} +{\rm
D}\Big(\frac{r_1+r_{12}}{2}g\Big)\right]-{\rm D}\left(\frac{\ln
r_{12}}{r_1}\right)-\frac{3}{2}{\rm D}\left(\frac{\ln
r_1}{r_{12}}\right)\nonumber\\&-&\frac{r_2}{2r_1^2r_{12}^2}
+\frac{1}{2r_1^2r_{12}}-\frac{1}{2r_1r_{12}^2}\;,\\
K_1 &=& {\rm D}\left(\frac{1}{r_2}\ln\Big[\frac{r_1}{r_{12}}\Big]\right)
-\frac{1}{2r_1^2r_2}+\frac{1}{2r_2r_{12}^2}+\frac{r_2}{2r_1^2r_{12}^2}\;.
\end{eqnarray}\end{subequations}
Anyway, we find that the exact solution of the linearized perturbation
equation is

\begin{eqnarray}\label{14}
\sigma_{(1)}&=&\frac{1}{16}{\rm D}\left(\frac{g}{3}
+\frac{r_1+r_2}{2r_{12}}\right)\nonumber\\
&+&\alpha_1\left\{\frac{1}{2}\Delta_1\left[\frac{g}{r_{12}} +{\rm
D}\left(\frac{r_1+r_{12}}{2}g\right)\right] -\frac{1}{2}{\rm
D}\left(\frac{\ln r_{12}}{r_1}\right)-\frac{15}{32}{\rm
D}\left(\frac{\ln r_1}{r_{12}}\right)+\frac{{\rm
D}g}{12r_1}\right.\nonumber\\ &&\quad\left.
-\frac{1}{32}\Delta_2\left(\frac{g}{r_{12}}\right)
-\frac{15r_2}{64r_1^2r_{12}^2}+\frac{1}{4r_1^2r_{12}}-\frac{1}{64r_1^2r_2}
-\frac{7}{24r_1r_{12}^2}+\frac{3}{64r_2r_{12}^2}\right\}\nonumber\\
&+&\alpha_2\Big\{1\leftrightarrow 2\Big\}\;.
\end{eqnarray}
Let us quote also, for completeness, the fully explicit form obtained
by expanding all the derivatives in this result:

\begin{eqnarray}\label{140}
\sigma_{(1)} &=& \frac{1}{96r_1r_2}+\frac{r_1+r_2}{64r_{12}^3}
+\frac{1}{192r_{12}}\left(\frac{1}{r_1}+\frac{1}{r_2}\right)
-\frac{1}{64r_{12}^3}\left(\frac{r_1^2}{r_2}
+\frac{r_2^2}{r_1}\right)\nonumber\\
&+&\alpha_1\left\{-\frac{1}{4r_1^3}-\frac{13}{64r_{12}^3}
-\frac{1}{24r_1r_{12}^2}-\frac{5}{192r_1^2r_{12}}+\frac{5}{192r_1^2r_2}
-\frac{r_1}{32r_2r_{12}^3}+\frac{3}{64r_2r_{12}^2} \right.
\nonumber\\ &&~\quad
\left. -\frac{1}{24r_1r_2r_{12}}+\frac{r_2}{4r_1r_{12}^3}
-\frac{15r_2}{64r_1^2r_{12}^2}+\frac{r_2}{4r_1^3r_{12}}
+\frac{15r_2^2}{64r_1^2r_{12}^3} +\frac{r_2^2}{4r_1^3r_{12}^2}
-\frac{r_2^3}{4r_1^3r_{12}^3}\right\}\nonumber\\ &+&\alpha_2\Big\{
1\leftrightarrow 2\Big\}\;.
\end{eqnarray}

The terms in the first line of Eqs. (\ref{14}) or (\ref{140})
contribute at the 2PN order in the conformal factor and they are in
agreement with Eq. (\ref{11''}). The other terms, proportional to
$\alpha_1$ or $\alpha_2$, will not contribute before the 3PN order
[because $\alpha_1$ and $\alpha_2$ carry a factor $1/c^2$ in
front]. Therefore, only a small part of the linearized approximation,
the one given by the first term in (\ref{14}), is necessary in the
proof of Theorem \ref{th1}. As a matter of fact, the non-linear
perturbation we consider contains much more information than a mere
post-Newtonian (2PN) expansion. Theorem \ref{th1} does not constitute
a very stringent requirement on the non-linear solution of the
perturbation equations.

A point we make by writing the linear solution $\sigma_{(1)}$ into the
primary form (\ref{13'}), involving the intermediate functions $H_1$
and $K_1$, is that the latter functions do enter in the post-Newtonian
metric at the 3PN order --- with the same numerical coefficients as
predicted by Eq. (\ref{13'}). This can be inferred from the expression
of the 3PN spatial metric given by Eq. (111) in Ref. \cite{livrev},
which contains the particular non-linear potential called $\hat{X}$,
together with the way that the potential $\hat{X}$ contains the
functions $H_1$ and $K_1$ as calculated by Eq. (6.11) in
\cite{BFP98}. Thus, although our solution agrees with post-Newtonian
calculations up to the 2PN order only, it does contain some correct
3PN features.

\subsection{Quadratic and higher-order approximations}\label{sec3C}

At the level of the second-order perturbation ($\propto\beta^2$) the
equation reads

\begin{eqnarray}\label{14'}
\Delta\sigma_{(2)}&=&\frac{1}{8}\sigma_{(1)}\partial_{ij}s_{ij}
+\partial_js_{ij}\partial_i\sigma_{(1)}
+s_{ij}\partial_{ij}\sigma_{(1)}\nonumber\\
&+&\frac{1}{8}\bigg(s_{ij}\Delta
s_{ij}-2s_{ij}\partial_i\partial_ks_{jk}-\partial_js_{ij}\partial_ks_{ik}
+\frac{3}{4}\partial_ks_{ij}\partial_ks_{ij}
-\frac{1}{2}\partial_ks_{ij}\partial_is_{jk}\bigg)\psi\nonumber\\
&+&\bigg(-s_{ij}\partial_ks_{jk}-s_{jk}\partial_ks_{ij}
+\frac{1}{2}s_{jk}\partial_is_{jk}\bigg)\partial_i\psi
-s_{ik}s_{jk}\partial_{ij}\psi\;.
\end{eqnarray}
At the next level, cubic-order, the equation will be made of the same
terms as in Eq. (\ref{14'}) but with the replacements
$\sigma_{(1)}\rightarrow\sigma_{(2)}$ and
$\psi\rightarrow\sigma_{(1)}$, together with many other terms that are
purely cubic in $s_{ij}$. And so on for the higher-order equations.

We shall see in Section \ref{sec4} that in order to prove the
agreement with the 2PN binary's {\it energy} (in contrast with the 2PN
{\it metric}), we need the full information content about the
linearized solution given by Eqs. (\ref{14})-(\ref{140}), and also a
crucial piece coming from the {\it second-order} perturbation,
solution of Eq. (\ref{14'}). In the case of non-linear perturbations
($n\geq 2$), it is in general impossible to find a solution in
analytic closed form. Fortunately, what we shall need is only to
control the expansion of $\sigma_{(2)}$ at spatial infinity (i.e. in
the far zone, $r\to +\infty$). And this {\it can} be achieved, without
disposing of the exact expression of $\sigma_{(2)}$, from the
knowledge of the far-zone expansion of the corresponding source-term
$\Sigma_{(2)}$. More generally, the far-zone expansion of the solution
$\sigma_{(n)}$ for any $n$ can be obtained on condition that the
far-zone expansion of its source $\Sigma_{(n)}$ has been determined
beforehand (say, by induction on $n$).

The method is issued from the investigation, in
Refs. \cite{B98mult,PB02}, of the multipole expansion of the solution
of a Poisson equation with non-compact-support source (i.e. whose
support is ${\mathbb R}^3$). In the present context, the multipole
expansion is completely equivalent to the far-zone expansion, when
$r\to +\infty$. Following Eq. (C.9) in Ref. \cite{PB02}, we obtain the
(formal) multipole expansion of $\sigma_{(n)}$ in the
form\,\footnote{Technical notations in Eq. (\ref{15}) are
$\Delta^{-1}$ for the standard Poisson integral; $L\equiv i_1\cdots
i_\ell$ for a multi-index composed of $\ell$ indices;
$\partial_L\equiv\partial_{i_1}\cdots \partial_{i_\ell}$ for the
product of $\ell$ partial derivatives; $y_L\equiv y_{i_1}\cdots
y_{i_\ell}$ for the product of $\ell$ spatial vectors. We do not write
the $\ell$ summation symbols over the $\ell$ indices composing $L$.}

\begin{eqnarray}\label{15}
{\cal M}(\sigma_{(n)})&=&FP\bigg\{\Delta^{-1}\left[{\cal
M}(\Sigma_{(n)})\right]-\frac{1}{4\pi}\sum_{\ell=0}^{+\infty}
\frac{(-)^\ell}{\ell!}
\partial_L\left(\frac{1}{r}\right)\int_{{\mathbb R}^3} d^3{\bf y}
\,y_L \Sigma_{(n)}({\bf y})\bigg\}\;,
\end{eqnarray}
where the calligraphic letter ${\cal M}$ refers to the multipole or
equivalently the far-zone expansion. The first term in the right side
of (\ref{15}) represents the effect of integrating ``term by term''
the multipole expansion of the source $\Sigma_{(n)}$ (this term exists
only in the case of non-compact-support sources, because the multipole
expansion of a compact-support function is zero). The second term in
Eq. (\ref{15}) is parametrized by the ``multipole moments'' associated
with the solution, which are given by some definite integrals
extending over the non-compact support ${\mathbb R}^3$ of
$\Sigma_{(n)}$. The symbol {\it FP} acts on both terms of (\ref{15}),
and stands for a certain operation of taking the {\it Finite Part},
defined in Refs. \cite{B98mult,PB02} by means of a process of complex
analytic continuation (with parameter $B\in {\mathbb C}$). The finite
part has proved to play a crucial role, in the case of
non-compact-support sources like $\Sigma_{(n)}$, in order to ensure
the well-definiteness of the integrals giving the multipole moments in
(\ref{15}). Notice that Eq. (\ref{15}) has been proved, in
Refs. \cite{B98mult,PB02}, to hold in the case of a regular source
[i.e. $C^\infty({\mathbb R}^3)$]. Nevertheless, in the presence of the
singular points ${\bf y}_1$ and ${\bf y}_2$, the formula (\ref{15})
can still be applied, but only in the case of that solution
$\sigma_{(n)}$ for which $\Delta\sigma_{(n)}=\Sigma_{(n)}$ is
satisfied in the sense of distributions.

\section{Total mass and individual black-hole masses}\label{sec4}

\subsection{Asymptotic structure of the solution}\label{sec4A}

A central result of the present paper concerns the ``asymptotics'' of
our solution --- both at spatial infinity and in the vicinity of the
two particles.

\begin{theorem}\label{th2}
(i) The metric $\gamma_{ij}$ is asymptotically flat at infinity (when
$r\equiv |{\bf x}|\to +\infty$), i.e.

\begin{equation}\label{16a}
\gamma_{ij} = \delta_{ij}+{\cal O}\left(\frac{1}{r}\right)\;.
\end{equation}
(ii) The two singular points ${\bf y}_1$ and ${\bf y}_2$ are the
images, via some appropriate inversions of the radial coordinates:
$\rho_1\sim 1/r_1$ and $\rho_2\sim 1/r_2$, of two ``asymptotically
finite'' regions (when $\rho_1\to +\infty$ and $\rho_2\to +\infty$),
in the sense that the metric, in coordinates $(\rho_1,{\bf n}_1)$,
behaves like

\begin{equation}\label{17a}
\Gamma_{ij}(\rho_1,{\bf n}_1)=\delta_{ij}
+\pi_{ij}({\bf n}_1)+{\cal
O}\left(\frac{1}{\rho_1}\right)\;,
\end{equation}
where $\pi_{ij}$ denotes a certain function of the angles.
\end{theorem}

Theorem \ref{th2} says that the solution is composed of an
asymptotically flat universe connected by continuity, {\it via} some
Einstein-Rosen-like bridges, to two other regions which are
asymptotically finite in the sense of Eq. (\ref{17a}). Note that,
though the metric (\ref{17a}), unlike the corresponding
Brill-Lindquist metric, is not asymptotically flat in the vicinity of
the two particles ({\it stricto-sensu}), the violation of asymptotic
flatness is not very severe, because the metric tends toward a
constant with respect to $\rho_1$, and does not involve any divergency
``at infinity'': it remains asymptotically finite --- hence the
name. On the other hand, what is very important is that the {\it real}
universe, as depicted by this solution, {\it is} asymptotically flat
at spatial infinity in the usual sense of Eq. (\ref{16a}). The global
structure described by Theorem \ref{th2} represents an attractive
feature, we argue, for considering the solution as a physically
well-motivated initial state of binary black holes.

The asymptotic flatness when $r\to +\infty$ follows from the far-zone
expansion of the generating function (\ref{9b}), which is easily
checked to start at $s_{ij}={\cal O}\left(\frac{1}{r}\right)$. Using
this fact we find that the source term for the linearized perturbation
--- i.e. the right side of Eq. (\ref{13}) --- behaves like
$\Sigma_{(1)}={\cal O}\left(\frac{1}{r^3}\right)$. With the help of
Eq. (\ref{15}) we readily obtain $\sigma_{(1)}={\cal
O}\left(\frac{1}{r}\right)$, as can also be checked directly with our
exact results (\ref{14})-(\ref{140}). Then, similarly, we deduce, by
induction on the non-linear order $n$ [using Eq. (\ref{15}) at each
step], that $\sigma_{(n)}={\cal O}\left(\frac{1}{r}\right)$ for any
$n$. So $\sigma={\cal O}\left(\frac{1}{r}\right)$, and the result
follows.

The main point about Theorem \ref{th2} is the behaviour of the
solution in the vicinity of the two particles. When $r_1\to 0$, we
find that $s_{ij}$ admits a {\it bounded} expansion [i.e. $s_{ij}$
does not blow up when $r_1\to 0$], of the type

\begin{equation}\label{18}
\beta s_{ij}=\varepsilon_{ij}({\bf n}_1)+{\cal O}(r_1)\;,
\end{equation}
where $\varepsilon_{ij}$ is a function of ${\bf n}_1=\frac{{\bf
x}-{\bf y}_1}{r_1}$, the direction of approach to the singularity
[$\varepsilon_{ij}$ is given by Eq. (\ref{a2a})]. From Eq. (\ref{18}),
the linearized source-term $\Sigma_{(1)}$ when $r_1\to 0$ is like some
$\Sigma_{(1)}={\cal O}\left(\frac{1}{r_1^3}\right)$. Now by
integrating ``term by term'' the expansion of $\Sigma_{(1)}$, we get a
similar expansion, but which starts at the order ${\cal
O}\left(\frac{1}{r_1}\right)$. Clearly, the solution $\sigma_{(1)}$,
when $r_1\to 0$, will be composed of the latter expansion, which
represents a particular solution, and augmented by a possible {\it
homogeneous} solution, solving the (source-free) Laplace
equation. However, because our original Poisson equation
$\Delta\sigma_{(1)}=\Sigma_{(1)}$ is satisfied in the sense of
distributions, the only possible homogeneous solution we can add is
one of the type ``regular at the origin'', taking the form of a sum of
STF products ${\hat x}_L\equiv {\rm STF}(x_{i_1}\cdots
x_{i_\ell})$. Because it is regular when $r_1\to 0$, this homogeneous
solution does not modify the leading singular behaviour of
$\sigma_{(1)}$, that we have therefore proved to be given by
$\sigma_{(1)}={\cal O}\left(\frac{1}{r_1}\right)$. And, again, this
can be checked with Eqs. (\ref{14})-(\ref{140}). The argument is
easily generalized to any non-linear order $n$ (by induction on $n$),
thus we conclude that the non-linear perturbation $\sigma$ diverges
when $r_1\to 0$, but not faster than $\sigma={\cal
O}(\frac{1}{r_1})$. As a result, the expansion of the conformal factor
$\Psi=\psi+\sigma$ is of the type

\begin{equation}\label{180'}
\Psi=\frac{\zeta({\bf
n}_1)}{r_1}+{\cal O}\left(r_1^0\right)\;,
\end{equation}
where $\zeta$ depends on the unit direction ${\bf n}_1$ but not on
$r_1$. From this we deduce that
$\gamma_{ij}=\Psi^4\widetilde{\gamma}_{ij}$ behaves dominantly when
$r_1\to 0$ like

\begin{equation}\label{18'}
\gamma_{ij}({\bf x})=\left(\frac{\zeta({\bf
n}_1)}{r_1}\right)^4\Big[\delta_{ij}+\varepsilon_{ij}({\bf
n}_1)+{\cal O}\left(r_1\right)\Big]\;.
\end{equation}
See Eq. (\ref{a3a}) for the expression of $\zeta({\bf n}_1)$ at
linearized order.

Let us now perform some inversion of the radial coordinate $r_1$. We
consider the coordinate change, valid in a neighbourhood of the
particle 1, that is defined by ${\bf x}\rightarrow (\rho_1,{\bf
n}_1)$, where

\begin{equation}\label{18''}
\rho_1=\frac{\zeta^2({\bf n}_1)}{r_1}\;.
\end{equation}
First we have $d x^i=d r_1^i$, where $r_1^i\equiv (r_1,{\bf n}_1)$,
because the particle is at rest: $v_1^i=0$. Then, the coordinate
transformation $r_1^i\rightarrow \rho_1^i\equiv (\rho_1,{\bf n}_1)$
involves simply the change of the radial variable given by
Eq. (\ref{18''}), with the angular part ${\bf n}_1$ being
unchanged\,\footnote{More general coordinate transformations,
involving a change of the unit direction ${\bf n}_1$, are also
possible.}. We compute

\begin{subequations}\label{180''}\begin{eqnarray}
d r_1^i&=&\left(\frac{\zeta}{\rho_1}\right)^2\left(\delta^{ij}
-2n_1^in_1^j+2n_1^i\chi^j\right)d \rho_1^j\;,\label{180a}\\
\chi^j&\equiv&\left(\delta^{jk}
-n_1^jn_1^k\right)\frac{\partial\ln\zeta}{\partial
n_1^k}\;.\label{180b}
\end{eqnarray}\end{subequations}
(Notice that $n_1^j\chi^j=0$.) We then find that, in the new
coordinate system $\rho_1^i=(\rho_1,{\bf n}_1)$, the metric, say
$\Gamma_{ij}$, admits when $\rho_1\to +\infty$ the {\it bounded}
expansion announced in Eq. (\ref{17a}), in which the quantity
$\pi_{ij}({\bf n}_1)$ reads

\begin{equation}\label{19'}
\pi_{ij}\equiv -4n_1^{(i}\chi^{j)}+4\chi^i\chi^j+
(\delta^{ki} -2n_1^kn_1^i+2n_1^k\chi^i)
(\delta^{lj} -2n_1^ln_1^j+2n_1^l\chi^j)\varepsilon_{kl}\;.
\end{equation}
It is such that $n_1^in_1^j\varepsilon_{ij}=n_1^in_1^j\pi_{ij}$. [See
also Eqs. (\ref{a4b}) and (\ref{a5}).] This completes the proof of
Theorem \ref{th2}.

\subsection{The ADM mass}\label{sec4B}

Because the metric is asymptotically flat at spatial infinity (Theorem
\ref{th2}), the binary's total ADM mass is given by the usual surface
integral on a topological 2-sphere at infinity:

\begin{equation}\label{16}
M=\frac{c^2}{16\pi G}\,\lim_{r\to +\infty}\int
dS^i\Big(\partial_j\gamma_{ij} -\partial_i\gamma_{jj}\Big)\;,
\end{equation}
where $dS^i$ is the outward surface element on the surface at infinity
($dS^i=d\Omega r^2n^i$ in the case of a coordinate sphere). To be more
explicit about $M$, we recall from Section \ref{sec4A} that some
functions $A_{ij}$ and $X$ of the unit direction ${\bf n}={\bf x}/r$
exist so that, when $r\to +\infty$,

\begin{subequations}\label{400}\begin{eqnarray}
\beta s_{ij}&=&\frac{A_{ij}({\bf n})}{r}+{\cal
O}\left(\frac{1}{r^2}\right)\;,\label{400a}\\ \sigma&=&\frac{X({\bf
n})}{r}+{\cal O}\left(\frac{1}{r^2}\right)\;.\label{400b}
\end{eqnarray}\end{subequations}
Here $A_{ij}$ is simply linear in $\beta$, while $X$ is in the form of
a full non-linear series in $\beta$: $X=\beta X_{(1)}+\beta^2
X_{(2)}+\cdots$ (see also the Appendix). In terms of these functions
$M$ is given by

\begin{equation}\label{401}
M=\frac{c^2}{G}\left\{2(\alpha_1+\alpha_2)
+\int\frac{d\Omega}{4\pi}\left[2X+\frac{1}{2}n^in^jA_{ij}\right]\right\}\;.
\end{equation}
The first term is identical to the result in the Brill-Lindquist
solution.

The computation of Eq. (\ref{401}) is quite straightforward at the
level of the {\it linearized} approximation, thanks to the closed-form
expression (\ref{14}). However, our aim will be the computation of the
2PN energy in Section \ref{sec5}, and we can ascertain beforehand (by
counting the required powers of $G$), that among all the terms
contributing to the 2PN energy there must be one coming from the {\it
second-order} perturbation, solution of Eq. (\ref{14'}). We find that
this particular non-linear term is to be computed only in $M$ (not in
the individual masses $m_1,\,m_2$). It comes from that part of
$\sigma_{(2)}$ --- or, more precisely, of its leading order
coefficient $X_{(2)}$ (when $r\to +\infty$) --- which does not involve
the constants $\alpha_1$ and $\alpha_2$, i.e. the part which would be
the analogue of the first term in $\sigma_{(1)}$ as given by
Eq. (\ref{14}). The other parts, proportional to $\alpha_1$ or
$\alpha_2$, appear at the 3PN order at least.

We succeeded in obtaining this non-linear term thanks to the method
described by Eq. (\ref{15}). First, one can check that the source-term
behaves like $\Sigma_{(2)}={\cal O}\left(\frac{1}{r^4}\right)$, so
there is no direct contribution, computed ``term-by-term'' from the
expansion of the source-term, to the multipole expansion at the order
${\cal O}\left(1/r\right)$: i.e. the first term in the right side of
(\ref{15}) is at least ${\cal O}\left(1/r^2\right)$. The looked-for
non-linear contribution is therefore given directly by the value of
the ``mass monopole'' of $\sigma_{(2)}$, i.e. the integral appearing
in the second term of Eq. (\ref{15}) for $\ell=0$. So,

\begin{equation}\label{200'}
\sigma_{(2)}=-\frac{1}{4\pi r}\int_{{\mathbb R}^3} d^3{\bf
y}\Sigma_{(2)}({\bf y})+{\cal O}\left(\frac{1}{r^2}\right)\;.
\end{equation}
The integral (\ref{200'}) can be computed analytically\,\footnote{This
integral is convergent, thus it is unnecessary to include a
finite-part operation {\it FP} (for notational simplicity we skip the
multipole-expansion symbol ${\cal M}$).}, and yields the following
contribution to the ADM mass:

\begin{equation}\label{20'}
2X_{(2)}=-\frac{1}{2\pi}\int_{{\mathbb R}^3} d^3{\bf
y}\Sigma_{(2)}({\bf y})=\frac{1}{128r_{12}^3}+{\cal
O}\left(\alpha\right)\;,
\end{equation}
where the remainder ${\cal O}\left(\alpha\right)$ indicates that the
terms proportional to $\alpha_1$ or $\alpha_2$ (in this second-order
perturbation $\propto\beta^2$), are not to be considered for the
present calculation. We thereby obtain $M$ with sufficient accuracy
for controlling the energy at the 2PN order\,\footnote{This means, by
the way, that the relative accuracy on $M$ itself is actually 3PN ---
because of the rest-mass contribution.}. The result reads

\begin{equation}\label{21}
M=\frac{c^2}{G}\left\{2(\alpha_1+\alpha_2)+\frac{\beta(\alpha_1
+\alpha_2)}{24r_{12}^2}
+\frac{\beta^2}{r_{12}^3}\left[\frac{1}{128}+{\cal
O}\left(\alpha\right)\right]+{\cal O}\left(\beta^3\right)\right\}\;.
\end{equation}
The cubic and higher-order perturbations ${\cal
O}\left(\beta^3\right)$ are neglected. See the Appendix for the
explicit expansion coefficients needed in this computation.

\subsection{The black-hole masses}\label{sec4C}

The situation as concerns the two individual masses $m_1$ and $m_2$ is
less easy than with the ADM mass because we dispose only of the notion
of ``asymptotic finiteness'' in the vicinity of the particles,
described by the fall-off property (\ref{17a}). Nevertheless, we wish
to find an appropriate concept for the black-hole masses. What we
shall do is to {\it define} $m_1$ by the same formula as for the ADM
mass $M$, but using the coordinate system $\rho_1^i=(\rho_1,{\bf
n}_1)$ in the limit where $\rho_1\to +\infty$. Accordingly we pose

\begin{equation}\label{17}
m_1=\frac{c^2}{16\pi G}\,\lim_{\rho_1\to +\infty}\int
dS_1^i\bigg(\frac{\partial\Gamma_{ij}}{\partial \rho_1^j}
-\frac{\partial\Gamma_{jj}}{\partial \rho_1^i}\bigg)\quad\hbox{and
$~1\leftrightarrow 2$}\;,
\end{equation}
where $\Gamma_{ij}(\rho_1,{\bf n}_1)$ denotes the metric in the
coordinates $\rho_1^i$, with $\rho_1$ being shown in Eq. (\ref{18''}),
and where the outward surface element reads
$dS_1^i=d\Omega_1\rho_1^2n_1^i$ in the case of a coordinate sphere.

Because $\Gamma_{ij}$ is not asymptotically flat in the usual sense,
due to the term with $\pi_{ij}$ in Eq. (\ref{17a}), the mass defined
by the previous integral will be typically unbounded (i.e. $m_1$ will
in general tend toward infinity like $\rho_1$). Therefore, the
definition (\ref{17}) does not {\it a priori} make sense. A
possibility would be to discard the divergent part ($\propto\rho_1$)
of the mass, and thereby to consider only the finite part of the
integral in Eq. (\ref{17}), i.e. the coefficient of the zeroth power
of $\rho_1$ in the expansion at infinity. This could represent an
appropriate postulate for the mass in a general situation [the finite
part prescription would be similar to the {\it FP} present in
Eq. (\ref{15})]. However such a ``finite part'' process imposed into
the definition of the mass appears to represent (untill more
convincing justification is proposed) a somewhat artificial and {\it
ad-hoc} recipe\,\footnote{The ADM mass, given by the surface term that
arises in the Hamiltonian, has been obtained even for space-times that
are not asymptotically flat \cite{HH95}. However, it seems difficult
to apply this result in the present case, notably because of our lack
of knowledge of the lapse function N in the vicinity of the particles
(after radial inversion $r_1\rightarrow\rho_1$ of the
coordinates).}. Gladly enough, we shall not need to invoke any {\it
ad-hoc} finite part because we shall prove that if we restrict
ourselves to the computation of the binary's energy at the 2PN order,
which represents anyway the maximal order at which our solution is
physically relevant, the masses $m_1$ and $m_2$ needed in this
computation {\it are} perfectly well-defined.

To compute $m_1$ and $m_2$, we must control the expansion of the
metric when $r_1\to 0$ to one order beyond
Eqs. (\ref{18})-(\ref{180'}). Let us pose

\begin{subequations}\label{402}\begin{eqnarray}
\beta s_{ij}&=&\varepsilon_{ij}({\bf n}_1)+r_1\mu_{ij}({\bf
n}_1)+{\cal O}(r_1^2)\;,\label{402a}\\ \Psi&=&\frac{\zeta({\bf
n}_1)}{r_1}+\eta({\bf n}_1)+{\cal O}\left(r_1\right)\;.\label{402b}
\end{eqnarray}\end{subequations}
[The expressions of the coefficients are relagated to
Eqs. (\ref{a2})-(\ref{a3}) in the Appendix.] It is straightforward to
derive from this the expression of the metric $\Gamma_{ij}$ in the
coordinate system $(\rho_1,{\bf n}_1)$, taking into account the term
$\sim 1/\rho_1$ in Eq. (\ref{17a}):

\begin{equation}\label{402'}
\Gamma_{ij}=\left(1+4\frac{\zeta\eta}{\rho_1}\right)\big[\delta_{ij}
+\pi_{ij}\big]+\frac{\zeta^2}{\rho_1}\kappa_{ij}+{\cal
O}\left(\frac{1}{\rho_1^2}\right)\;,
\end{equation}
where $\kappa_{ij}=(\delta^{ki}-2n_1^kn_1^i+2n_1^k\chi^i)(\delta^{lj}
-2n_1^ln_1^j+2n_1^l\chi^j)\mu_{kl}$. Then, by inserting it into
Eq. (\ref{17}), we obtain

\begin{equation}\label{403}
m_1=\frac{c^2}{G}\lim_{\rho_1\to
+\infty}\int\frac{d\Omega_1}{4\pi}\left[-\frac{1}{4}
\rho_1\pi_{jj}+2\eta\zeta\big(1+n_1^in_1^j\varepsilon_{ij}\big)
+\frac{1}{2}\zeta^2n_1^in_1^j\mu_{ij}\right]\;.
\end{equation}
Here, $\pi_{jj}$ denotes the trace of the quantity (\ref{19'}), and we
have used the facts that
$n_1^in_1^j\pi_{ij}=n_1^in_1^j\varepsilon_{ij}$ and
$n_1^in_1^j\kappa_{ij}=n_1^in_1^j\mu_{ij}$, together with the useful
cancellation of the angular integral

\begin{equation}\label{404}
\int\frac{d\Omega_1}{4\pi}n_1^in_1^j\varepsilon_{ij}({\bf n}_1)=0\;,
\end{equation}
which can be checked from Eq. (\ref{a3a}) in the Appendix.

We observe that Eq. (\ref{403}) contains a term proportional to
$\rho_1$, with $\pi_{jj}$ as a factor, which can and {\it does} make
the mass to become infinite in the limit $\rho_1\to +\infty$ [indeed,
see Eq. (\ref{a6})]. As suggested above, such an infinite term could
be removed by some procedure of taking the finite part (whose meaning
would {\it a priori} be quite unclear), but we find that this infinite
term is in fact ``negligible'' for the present purpose, because
$\pi_{jj}$ starts only at non-linear order in $\beta$,

\begin{equation}\label{405}
\pi_{jj}=4\chi^j\chi^j+4
\big(n_1^k\chi^l+n_1^kn_1^l\chi^j\chi^j\big)\varepsilon_{kl}={\cal
O}\left(\beta^2\right)\;.
\end{equation}
[See also the expression (\ref{a5}).] Therefore we get, at the
linearized level, a finite expression for the mass (formally):

\begin{equation}\label{406}
m_1=\frac{c^2}{G}\int\frac{d\Omega_1}{4\pi}\left[
2\eta\zeta\big(1+n_1^in_1^j\varepsilon_{ij}\big)
+\frac{1}{2}\zeta^2n_1^in_1^j\mu_{ij}\right]+{\cal
O}\left(\beta^2\right)\;.
\end{equation}

This expression is sufficient to control the 2PN energy. The
non-linear corrections ${\cal O}\left(\beta^2\right)$ are infinite but
will not be needed in this paper. In a sense, they do not belong to
the ``realm'' of the present solution, which is limited to the physics
at the 2PN approximation: i.e. agreement with the 2PN metric in the
near-zone, and internal consistency of the asymptotics up to the 2PN
order as regards the energy content of the solution (see Section
\ref{sec5})\,\footnote{Recall that our proposal for binary black-hole
initial data is to adopt the conformal metric (\ref{3}) and to deduce
the conformal factor $\Psi$ from it by {\it numerical} methods. The
present {\it analytic} investigation [including the theoretical
definition of the mass (\ref{17})] is aimed at verifying the physical
soundness of this proposal.}. As for the linearized terms in
Eq. (\ref{406}), they are dealt with thanks to the explicit result
(\ref{14}), and we get

\begin{equation}\label{22}
m_1=\frac{2c^2\alpha_1}{G}\left[1+\frac{\alpha_2}{r_{12}}
+\frac{\beta}{48r_{12}^2}\left(1-\frac{25\alpha_2}{r_{12}}\right)
+{\cal O}\left(\beta^2\right)\right]\quad\hbox{and $~1\leftrightarrow
2$}\;.
\end{equation}
By setting $\beta=0$ into Eq. (\ref{22}), we reproduce the result
valid in the case of the Brill-Lindquist solution.

\section{The binary's geometrostatic energy}\label{sec5}

With the masses being defined and computed ``geometrically'' in
Section \ref{sec4}, we get the opportunity of an interesting
consistency check of our solution, concerning the ``geometrostatic''
energy that is associated with those masses. Indeed, in Theorem
\ref{th1} we recovered the 2PN metric in the near zone, therefore we
know the coordinate transformation $\delta x^i = \xi^i$ between our
presently used coordinate system and the harmonic one. The gauge
transformation vector has been obtained in Eq. (\ref{7}). Therefore it
is possible to control, without ambiguity --- because we determined
the coordinate transformation ---, the binary's interacting energy $E$
up to the same 2PN order. If our solution has anything cogent
(physically speaking), the latter energy should be in complete
agreement with the post-Newtonian result at the 2PN order known from
Refs. \cite{DD81,D82,B96,BFeom}.

\begin{theorem}\label{th3} 
The binary's interacting energy, deduced from those masses ${\rm
(\ref{16})}$ and ${\rm (\ref{17})}$ computed by surface integrals at
infinity, i.e.

\begin{equation}\label{201}
\frac{E}{c^2}=M-m_1-m_2\;,
\end{equation}
gives back the energy calculated by standard post-Newtonian methods at
the ${\rm 2PN}$ order (after invoking the same coordinate
transformation as in Theorem \ref{th1}).
\end{theorem}

The ADM mass has been previously obtained in the form (\ref{21}),
while the masses $m_1$ and $m_2$ follow from the result
(\ref{22}). The corresponding energy $E$ reads then

\begin{equation}\label{23}
E=\frac{c^4}{G}\left\{-\frac{4\alpha_1\alpha_2}{r_{12}}\left(1
-\frac{25\beta}{48r_{12}^2}\right)
+\frac{\beta^2}{r_{12}^3}\left[\frac{1}{128}+{\cal
O}\left(\alpha\right)\right]+{\cal O}\left(\beta^3\right)\right\}\;.
\end{equation}
As we said above, this formula is accurate enough to get full control
of the 2PN approximation. We also remarked that at this order the
energy is well-defined (i.e. finite).

Our task is to expand that energy when $c\to +\infty$, using the facts
that $\alpha_1,\,\alpha_2={\cal O}\left(\frac{1}{c^2}\right)$ and
$\beta={\cal O}\left(\frac{1}{c^4}\right)$ [actually, we already
applied the post-Newtonian approximation when arguing that the
remainder ${\cal O}\left(\alpha\right)$ is negligible]. Also, we know
that $\beta$ is given by Eq. (\ref{9a}). We first invert
Eq. (\ref{22}), in an iterative post-Newtonian way, so as to obtain
$\alpha_1,\,\alpha_2$ with relative 2PN accuracy. The result is

\begin{equation}\label{24}
\alpha_1=\frac{G m_1}{2c^2}\left[1-\frac{G m_2}{2r_{12}c^2}
+\frac{G^2m_2(5m_1+3m_2)}{12r_{12}^2c^4}+{\cal
O}\left(\frac{1}{c^6}\right)\right]\quad\hbox{and $~1\leftrightarrow
2$}\;.
\end{equation}
At the 1PN order, we are consistent with the Brill-Lindquist
prediction and with what is given by Eq. (\ref{11'}). But, quite
naturally, we find that Eq. (\ref{24}) differs from the
Brill-Lindquist result at the 2PN order\,\footnote{In the case of the
Brill-Lindquist solution we have
$$\alpha_1^{\rm BL}=\frac{G m_1}{2c^2}\left[1-\frac{G m_2}{2r_{12}c^2}
+\frac{G^2m_2(m_1+m_2)}{4r_{12}^2c^4}+{\cal
O}\left(\frac{1}{c^6}\right)\right]\;.$$}. 

Substituting (\ref{24}) back into (\ref{23}), we then arrive, after
suitable post-Newtonian expansion, at the expression of the 2PN energy
in terms of the two ``physical'' individual masses:

\begin{equation}\label{25}
E=-\frac{G m_1m_2}{r_{12}}+\frac{G^2m_1m_2(m_1+m_2)}{2r_{12}^2c^2}
-\frac{G^3m_1m_2(m_1^2+19m_1m_2+m_2^2)}{4r_{12}^3c^4}+{\cal
O}\left(\frac{1}{c^6}\right)\;.
\end{equation}
It is composed of the standard Newtonian potential energy (for static
black holes), augmented by 1PN and 2PN corrections. The result,
however, is not yet the one we want to prove, because the particles'
separation $r_{12}$ corresponds to our particular coordinate
system. If we want to compare $E$ with the post-Newtonian prediction,
we must take into account the coordinate transformation (in the ``near
zone'') that was determined in Theorem \ref{th1}.

The required link between $r_{12}$ and the particles' separation
$R_{12}$ in harmonic coordinates is readily obtained with the help of
Eq. (\ref{7}), which permits to compute the shift of the world-lines
that is induced by the coordinate transformation\,\footnote{We discard
an infinite ``self-term'' when considering the
coordinate-transformation vector $\xi_i({\bf x})$ at the singular
location of the two particles. Thus, from Eq. (\ref{7}), which is
valid {\it a priori} only outside the singularities, we obtain the
shift vector
$$ \xi_i({\bf y}_1)=\frac{G^2m_2^2}{4c^4}\frac{n_{12}^i}{r_{12}}+{\cal
O}\left(\frac{1}{c^6}\right)\;.$$ In other words we consider only the
finite part of $\xi_i({\bf x})$ when ${\bf x}\to {\bf y}_1$ (in
Hadamard's sense).}. We get

\begin{equation}\label{26}
r_{12}=R_{12}-\frac{G^2(m_1^2+m_2^2)}{4R_{12}c^4}+{\cal
O}\left(\frac{1}{c^6}\right)\;.
\end{equation}
The result we find after replacing Eq. (\ref{26}) into Eq. (\ref{25}),
and effecting the post-Newtonian re-expansion [the replacement is to
be made only into the Newtonian term of (\ref{25})], is

\begin{equation}\label{27}
E=-\frac{G m_1m_2}{R_{12}}+\frac{G^2m_1m_2(m_1+m_2)}{2R_{12}^2c^2}
-\frac{G^3m_1m_2(2m_1^2+19m_1m_2+2m_2^2)}{4R_{12}^3c^4}+{\cal
O}\left(\frac{1}{c^6}\right)\;.
\end{equation}
Most satisfactorily, we discover that (\ref{27}) is in complete
agreement with the prediction for the 2PN energy in harmonic
coordinates, calculated in Refs. \cite{DD81,D82,B96,BFeom}. The
expression we end up with is the same as given by Eq. (B6) in
Ref. \cite{B96} --- in which of course one must set the two particles'
velocities to zero.

The latter agreement is interesting, but we said that it is quite
mandatory if our solution makes sense. Technically speaking, it
necessitated the control of the metric up to second-order perturbation
theory on the Brill-Lindquist background [{\it cf.} the crucial
contribution of Eq. (\ref{20'}) to the ADM mass]. Furthermore, the
result (\ref{27}) can be said to check the global character of the
solution (notably the asymptotics therein) --- because the physical
masses have been computed by surface integrals at
infinity. Alternatively, it shows the relevance at 2PN order of the
definition (\ref{17}) we adopted for the black-hole individual masses.

Finally, let us comment that Theorem \ref{th3} tells us something
about the physical tenets at the basis of the usual post-Newtonian
approximation when it is applied to the description of black
holes. Indeed, the ``post-Newtonian'' masses $m_1$ and $m_2$, which
parametrize the post-Newtonian iteration, are introduced as being the
coefficients of Dirac delta-functions in the Newtonian density of
point-like particles \cite{livrev}. Now, we have just seen that in
fact these post-Newtonian masses agree (within the present approach at
least, i.e. in the time-symmetric situation, and up to the 2PN order)
with the ``geometrostatic'' masses which are associated with some
Einstein-Rosen-like bridges. We view this as a confirmation that the
post-Newtonian calculations, which treat formally the compact objects
by means of delta-function singularities\,\footnote{For instance, the
calculation of the 3PN equations of motion of compact binaries
reported in Ref. \cite{BFeom}.}, are appropriate to the description of
systems of black holes (as long as the {\it orbital} motion of the
black holes can be considered to be ``slow'' in the post-Newtonian
sense, i.e. in the so-called inspiralling phase of black-hole
binaries).

\acknowledgments

It is a pleasure to thank Gilles Esposito-Far\`ese and Eric
Gourgoulhon for discussions and remarks. This work was partially
supported by the International Academic Grant of the Hirosaki
University.

\appendix*

\section{Relevant expansion formulas}

In this Appendix we provide the explicit expressions of the various
coefficients in the expansion formulas introduced in Section
\ref{sec4} for the computation of the masses.

For the ADM mass we need the dominant term, when $r\to +\infty$, in
the generating function $s_{ij}$ and the solution of the linear-order
perturbation for the conformal factor [see Eqs. (\ref{400})]. They are
given by

\begin{subequations}\label{a1}\begin{eqnarray}
A_{ij}&=&\frac{\beta}{2r_{12}}n_{12}^{<i}n_{12}^{j>}\;,\label{a1a}\\
\beta X_{(1)}&=&\frac{\beta}{r_{12}}\left\{-\frac{1}{32}\left[({\bf
n}.{\bf n}_{12})^2-\frac{1}{3}
\right]+\frac{1}{48}\frac{\alpha_1+\alpha_2}{r_{12}}\right\}\;,\label{a1b}
\end{eqnarray}\end{subequations}
where ${\bf n}={\bf x}/r$ and ${\bf n}_{12}=({\bf y}_1-{\bf
y}_2)/r_{12}$ (with ${\bf n}.{\bf n}_{12}$ denoting the ordinary
scalar product). The second-order perturbation coefficient was already
computed in Eq. (\ref{20'}):

\begin{equation}\label{a1'}
\beta^2X_{(2)}=\frac{\beta^2}{r_{12}^3}\left[\frac{1}{256}+{\cal
O}\left(\alpha\right)\right]\;.
\end{equation}

For the masses $m_1$ and $m_2$ we require the expansions when $r_1\to
0$ of $s_{ij}$ and the conformal factor to one order beyond the
dominant term [see Eqs. (\ref{402})]. For $s_{ij}$ we have

\begin{subequations}\label{a2}\begin{eqnarray}
\varepsilon_{ij}&=&\frac{\beta}{r_{12}^2}\left(n_{12}^{<i}n_{12}^{j>}
-\frac{1}{2}n_1^{<i}n_{12}^{j>}\right)\;,\label{a2a}\\
\mu_{ij}&=&\frac{\beta}{r_{12}^3}\left[n_{12}^{<i}n_{12}^{j>}
\left(-{\bf n}_1.{\bf n}_{12}-\frac{3}{4}\right)
+n_1^{<i}n_{12}^{j>}\left(\frac{3}{4}{\bf n}_1.{\bf
n}_{12}+\frac{3}{4}\right)
-\frac{1}{4}n_1^{<i}n_1^{j>}\right]\;.\label{a2b}
\end{eqnarray}\end{subequations}
For the conformal factor:

\begin{subequations}\label{a3}\begin{eqnarray}
\zeta&=&\alpha_1\left[1+\frac{\beta}{r_{12}^2}\left(-\frac{1}{2}({\bf
n}_1.{\bf n}_{12})^2 +\frac{5}{24}{\bf n}_1.{\bf
n}_{12}+\frac{1}{6}\right)\right]+{\cal
O}\left(\beta^2\right)\;,\label{a3a}\\
\eta&=&1+\frac{\alpha_2}{r_{12}}+\frac{\beta}{r_{12}^2}\left[-\frac{1}{24}{\bf
n}_1.{\bf n}_{12}+\frac{1}{48}\right.\nonumber\\
&&\quad\qquad\qquad\left. +\frac{\alpha_1}{r_{12}}
\left(\frac{1}{4}({\bf n}_1.{\bf n}_{12})^3 +\frac{5}{32}({\bf
n}_1.{\bf n}_{12})^2 -\frac{5}{24}{\bf n}_1.{\bf
n}_{12}-\frac{5}{96}\right)\right.\nonumber\\
&&\quad\qquad\qquad\left. +\frac{\alpha_2}{r_{12}}\left(-\frac{1}{24}{\bf
n}_1.{\bf n}_{12} -\frac{25}{48}\right)\right]+{\cal
O}\left(\beta^2\right)\;.\label{a3b}
\end{eqnarray}\end{subequations}
The quantities $\chi^i$ and $\pi_{ij}$ defined by Eqs. (\ref{180b})
and (\ref{19'}) read (at linear order in $\beta$)

\begin{subequations}\label{a4}\begin{eqnarray}
\chi^i&=&\frac{\beta}{r_{12}^2}\left({\bf n}_1.{\bf
n}_{12}-\frac{5}{24}\right)\Big[-n_{12}^i+({\bf n}_1.{\bf
n}_{12})n_1^i\Big]+{\cal O}\left(\beta^2\right)\;,\label{a4a}\\
\pi_{ij}&=&\frac{\beta}{r_{12}^2}\left(n_{12}^{<i}n_{12}^{j>}
-\frac{1}{3}n_1^{<i}n_{12}^{j>}-\frac{1}{6}({\bf n}_1.{\bf
n}_{12})n_1^{<i}n_1^{j>}\right)+{\cal
O}\left(\beta^2\right)\;.\label{a4b}
\end{eqnarray}\end{subequations}
Note that at linear order $\pi_{ij}$ is trace-free. Its trace arises
only at second-order in $\beta$,

\begin{equation}\label{a5}
\pi_{jj}=\frac{\beta^2}{r_{12}^4}\left(-\frac{1}{6}{\bf n}_1.{\bf
n}_{12}+\frac{5}{144}\right)\Big[({\bf n}_1.{\bf
n}_{12})^2-1\Big]+{\cal O}\left(\beta^3\right)\;.
\end{equation}
At this order the angular average of $\pi_{ij}$ is non-zero:

\begin{equation}\label{a6}
\int\frac{d\Omega_1}{4\pi}\pi_{jj}=-\frac{5}{216}\frac{\beta^2}{r_{12}^4}+{\cal
O}\left(\beta^3\right)\;.
\end{equation}


\begin{thebibliography}{}
\bibitem{cook}For a review see: G.B. Cook, Living Rev. Relat. {\bf
3}, 5 (2000).
\bibitem{lichne}A. Lichnerowicz, J. Math. Pures Appl. {\bf 23}, 37 (1944).
\bibitem{brill}D.R. Brill, Ann. Phys. {\bf 7}, 466 (1959).
\bibitem{york}J.W. York, Phys. Rev. Lett. {\bf 26}(26), 1656 (1971);
and {ibid.} {\bf 28}(16), 1082 (1972).
\bibitem{misner}C.W. Misner, Ann. Phys. {\bf 24}, 102 (1963).
\bibitem{lindquist}R.W. Lindquist, J. Math. Phys. {\bf 4}, 938 (1963).
\bibitem{BL}D.R. Brill and R.W. Lindquist, Phys. Rev. {\bf 131}, 471
(1963).
\bibitem{S87}G. Sch\"afer, Phys. Lett. A {\bf 123}, 336 (1987).
\bibitem{DI91}T. Damour and B.R. Iyer, Ann. Inst. H. Poincar\'e {\bf
54}, 115 (1991).
\bibitem{BDI95}L. Blanchet, T. Damour and B.R. Iyer, Phys. Rev. D {\bf
51}, 5360 (1995).
\bibitem{WWi96}C.M. Will and A.G. Wiseman, Phys. Rev. D {\bf 54}, 4813
(1996).
\bibitem{BFP98}L. Blanchet, G. Faye and B. Ponsot, Phys. Rev. D{\bf 58},
124002 (1998).
\bibitem{AF97}H. Asada and T. Futamase, Prog. Th. Phys. Suppl. {\bf
128}, 123 (1997).
\bibitem{livrev}For a review see: L. Blanchet, Living Rev. Relat. {\bf
5}, 3 (2002).
\bibitem{alvi}K. Alvi, Phys. Rev. D{\bf 61}, 124013 (2000).
\bibitem{diener}W. Tichy, B. Br\"ugmann, M. Campanelli and P. Diener,
Phys. Rev. D{\bf 67}, 064008 (2003).
\bibitem{BCLT}J. Baker, M. Campanelli, C.O. Lousto and R. Takahashi,
Phys. Rev. D{\bf 65}, 124012 (2002).
\bibitem{matzner}R.A. Matzner, M.F. Huq and D. Shoemaker,
Phys. Rev. D{\bf 59}, 024015 (1999).
\bibitem{BMNM}E. Bonning, P. Marronetti, D. Neilsen and R.A. Matzner
(gr-qc/0305071).
\bibitem{GGB1}E. Gourgoulhon, P. Grandcl\'ement and S. Bonazzola,
Phys. Rev. D{\bf 65}, 044020 (2002).
\bibitem{GGB2}P. Grandcl\'ement, E. Gourgoulhon and S. Bonazzola,
Phys. Rev. D{\bf 65}, 044021 (2002).
\bibitem{B02ico}L. Blanchet, Phys. Rev. D {\bf 65}, 124009 (2002); see
also: L. Blanchet, in ``2001: a relativistic spacetime odyssey'',
eds. I. Ciufolini {et al.}, World Scientific, p. 411 (gr-qc/0209089).
\bibitem{DGG02}T. Damour, E. Gourgoulhon and P. Grandcl\'ement,
Phys. Rev. D{\bf 66}, 024007 (2002).
\bibitem{Smarr}L.L. Smarr, in ``Sources of gravitational radiation'',
ed. L.L. Smarr, Cambridge U. Press, Cambridge (1979).
\bibitem{AHSSS}P. Anninos, D. Hobill, E. Seidel, L. Smarr and
W.-M. Suen, Phys. Rev. Lett. {\bf 71}, 2851 (1993).
\bibitem{SPW95}L.E. Simone, E. Poisson and C.M. Will, Phys. Rev. D{\bf
52}, 4481 (1995).
\bibitem{B98mult}L. Blanchet, Class. Quantum Gravity {\bf 15}, 1971
(1998).
\bibitem{PB02}O. Poujade and L. Blanchet, Phys. Rev. D{\bf 65}, 124020
(2002).
\bibitem{HH95}S.W. Hawking and G.T. Horowitz, Class. Quantum Gravity
{\bf 13}, 1487 (1996).
\bibitem{DD81}T. Damour and N. Deruelle, C.R. Acad. Sc. Paris {\bf
293}, 877 (1981).
\bibitem{D82}T. Damour, C.R. Acad. Sc. Paris {\bf 294}, 1355 (1982).
\bibitem{B96}L. Blanchet, Phys. Rev. D{\bf 54}, 1417 (1996).
\bibitem{BFeom}L. Blanchet and G. Faye, Phys. Rev. D{\bf 63}, 062005
(2001).
\end{thebibliography}
\end{document}